\documentclass[aps,twocolumn,showpacs,preprintnumbers,superscriptaddress,amsmath,amssymb]{revtex4}
\usepackage{graphicx}
\usepackage{dcolumn}
\usepackage{bm}

\begin{document}

\title{In-plane substitution effect on the magnetic properties of two-dimensional Spin-gap system SrCu$_2$(BO$_3$)$_2$}

\author{G. T. liu}
\author{J. L. Luo}
\email{JLLuo@aphy.iphy.ac.cn} \affiliation{Beijing National
Laboratory of Condensed Matter Physics, Institute of Physics,
Chinese Academy of Sciences, Beijing 100080, People's Republic of
China}

\author{Y. Q. Guo}
\affiliation{Beijing National Laboratory of Condensed Matter
Physics, Institute of Physics, Chinese Academy of Sciences,
Beijing 100080, People's Republic of China}

\author{S. K. Su}
\affiliation{Beijing National Laboratory of Condensed Matter
Physics, Institute of Physics, Chinese Academy of Sciences,
Beijing 100080, People's Republic of China}

\author{P. Zheng}
\affiliation{Beijing National Laboratory of Condensed Matter
Physics, Institute of Physics, Chinese Academy of Sciences,
Beijing 100080, People's Republic of China}

\author{N. L. Wang}
\affiliation{Beijing National Laboratory of Condensed Matter
Physics, Institute of Physics, Chinese Academy of Sciences,
Beijing 100080, People's Republic of China}

\author{D. Jin}
\affiliation{Beijing National Laboratory of Condensed Matter
Physics, Institute of Physics, Chinese Academy of Sciences,
Beijing 100080, People's Republic of China}

\author{T. Xiang}
\affiliation{Institute of Theoretical Physics and
Interdisciplinary Center of Theoretical Studies, Chinese Academy
of Science, P.O. Box 2735, Beijing 100080, People's Republic of
China}

\date{\today}

\begin{abstract}
A series of in-plane substituted compounds, including Cu-site
(SrZn$_x$Cu$_{2-x}$(BO$_3$)$_2$), and B-site
(SrCu$_2$(Si$_x$B$_{1-x}$O$_3$)$_2$) substitution, were
synthesized by solid state reaction. X-ray diffraction
measurements reveal that these compounds are single-phase
materials and their in-plane lattice parameter depends
systematically on the substituting content $x$. The magnetic
susceptibility in different magnetic fields, the magnetization at
different temperatures, and the resistivity at room temperature
were measured, respectively. It is found that the spin gap deduced
from the magnetic susceptibility measurements decreases with
increasing of $x$ in both Cu- and B-site substitution. No
superconductivity was found in these substituted compounds.
\end{abstract}

 \maketitle

The psedudogap observed in the high-$\emph{T$_c$}$ cuprates has
stimulated intensified interest on the investigations of systems
with spin-gap. A novel spin-gap system discovered recently is
strontium copper borate SrCu$_2$(BO$_3$)$_2$\cite{Kageyama1}. This
material is a realization of the two-dimensional spin-$1/2$
Shastry-Sutherland model\cite{Shastry}. It has a tetragonal
structure constructed with alternating CuBO$_3$ and Sr
planes\cite{Smith}. At room temperature, the lattice parameters
are \emph{a}=8.995 \AA\ and \emph{c}=6.649 \AA\ . In CuBO$_3$
plane, all Cu$^{2+}$ ions with spin-$1/2$ are located at
crystallographically equivalent sites, and the nearest two
Cu$^{2+}$ ions form a dimer unit. The dimer units are connected
orthogonally through BO$_3$ triangles. The magnetic properties of
SrCu$_2$(BO$_3$)$_2$ can be effectively described by the
two-dimensional Heisnberg model:
\begin{equation}
\emph{H}=\emph{J}\sum_{NN}\emph{S}_i\emph{S}_j +
\emph{J$^\prime$}\sum_{NNN}\emph{S}_i\emph{S}_j
\end{equation}
where $\emph{J}$ and $\emph{J}^\prime$ are the intradimer and
interdimer exchange constants. The phase diagram of this
Shasty-Sutherland-type model has been
studied\cite{Koga,Zheng,Carpentier,Chung}. The ground state of the
model is gapped and described approximately by a product of local
dimer singlets when $J^\prime/J$$<$0.68\cite{Shastry, Miyahara}.
Recently experimental results for magnetic
susceptibility\cite{Kageyama1,Kageyama2}, Cu nuclear quadrupole
resonance\cite{Kageyama1}, high-field
magnetization\cite{Kageyama1}, electron-spin
resonance\cite{Nojiri}, Raman scattering\cite{Lemmens}, inelastic
neutron scattering\cite{Kageyama3, Kakurai}, and specific
heat\cite{Kageyama4} showed that the spin gap of this material is
$\sim$30 K. This material shows a plateaus in the magnetization
curve\cite{Kageyama1,Onizuka}. In order to explain this
phenomenon, great effort has been devoted to the study of
excitation state. It was found that the magnetization plateau is
associated with a superstructure induced by localized spin
excitations\cite{Kodama,Miyahara}. For SrCu$_2$(BO$_3$)$_2$, the
ratio $J^\prime/J$ is close to its critical value. The value of
$J^\prime/J$ can be changed by applying pressure or
substitution\cite{Leung,Marston}. On doping, this system may
select other ordered (or disordered) states so that electrons or
holes can move freely\cite{Marston}.

Recently it was suggested theoretically that this material can
become superconducting upon electron or hole
doping\cite{Leung,Chung,Kumar,Kimura}. A number of different
superconducting states were suggested at different doping
levels\cite{Kumar,Kimura,Chung}. To examine these theoretical
proposals, Kudo\cite{Kudo}, Lappas\cite{Lappas}, Choi\cite{Choi}
and Zorko\cite{Zorko1, Zorko2} $et$ $al$ investigated doping
effect of non-magnetic ions on SrCu$_2$(BO$_3$)$_2$. They found
that the spin gap is hardly to be suppressed by small substitution
of non-magnetic impurities. Our previous work\cite{Liu1} indicates
that substituting Sr can significantly reduce the spin gap and the
electrical resistivity, especially for La-doping. But the high
resistivity, in the order of $\sim$10$^8$ $\Omega$ m, indicating
that they are all insulators. These results suggested that these
doped carriers are all almost localized. The possible reason comes
from the B-O bond, which adopts $s$$p^2$ hybridization. The bond
is so stable that these doped carriers are localized in the
CuBO$_3$ plane. In this paper, we expect the in-plane substitution
might destroy or distort $sp^2$ hybridization and form some
conducting channels in CuBO$_3$, thus superconductivity maybe
occurs. Unfortunately, we found no such an effect. However, we
find that both Zn- and Si-substituting has pronounced effects on
the dimer structure and magnetic properties.

The samples of SrZn$_x$Cu$_{2-x}$(BO$_3$)$_2$ (Cu-site
substitution) and SrCu$_2$(Si$_x$B$_{1-x}$O$_3$)$_2$ (B-site
substitution) with different substituting content $x$ were
prepared by heating stoichiometric mixtures of SrCO$_3$, CuO,
B$_2$O$_3$, ZnO and SiO$_2$ with high purity. Initial chemicals
were mixed carefully, followed by heat treatment between 820 to
 $860^{\circ}$C in air for one week with several intermediate
grindings.

\begin{figure}[b]
\includegraphics[height=0.75\linewidth]{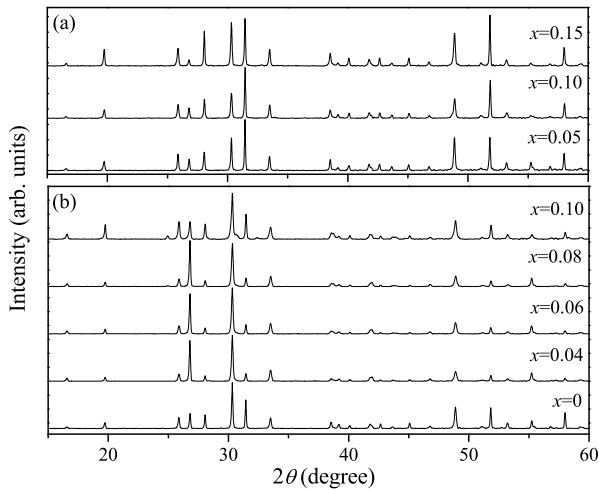}
\caption{\label{fig:fig_1} {The x-ray powder diffraction patterns
of (a) SrZn$_x$Cu$_{2-x}$(BO$_3$)$_2$ with \emph{x}$\leq$0.15, and (b)
SrCu$_2$(Si$_x$B$_{1-x}$O$_3$)$_2$ samples with
\emph{x}$\leq$0.1.}}
\end{figure}

\renewcommand\arraystretch{1.5}
\begin{table*}
\caption{\label{tab:Table1}Lattice parameters, $\chi_0$
($\times$$10^{-3}$ emu/(mol Cu)), $C^{'}$ ($\times$$10^{-3}$ emu
K/(mol Cu)), $\Delta$, resistivity $\rho$ ($\times$$10^8$ $\Omega$
m), effective moments $\mu_\textbf{eff}$, and the atomic ratio in
SrZn$_x$Cu$_{2-x}$(BO$_3$)$_2$ and
SrCu$_2$(Si$_x$B$_{1-x}$O$_3$)$_2$.}
\begin{ruledtabular}
\begin{tabular}{c|cccccccccccccc}
\ Formula &$x$&$a$(\AA )&$c$(\AA)  &$\chi_0$  &$C^{'}$ &$\Delta$(K) &$\rho$  &$\mu_\textbf{eff}$($\mu_B$) &Cu &B &Sr &Zn & Si\\
\hline
SrZn$_x$Cu$_{2-x}$(BO$_3$)$_2$ &0    & 8.995 & 6.649 &-0.740 &10.57 &21.5 &4.158 &0.110 & $$  & $$  & $$  & $$  & $$ & $$ \\
                               &0.05 & 8.990 & 6.648 &-0.898 &12.59 &20.1 &1.045 &0.156 &1.95 &1.97 &1.00 &0.05 & $$ & $$ \\
                               &0.10 & 8.989 & 6.649 &-1.043 &14.02 &16.7 &0.957 &0.193 &1.90 &1.93 &1.00 &0.09 & $$ & $$  \\
                               &0.15 & 8.984 & 6.647 &-2.889 &32.77 &15.0 &0.898 &0.212 &1.79 &1.95 &1.00 &0.11 & $$ & $$  \\
\hline
SrCu$_2$(Si$_x$B$_{1-x}$O$_3$)$_2$ &0.04 & 8.990 & 6.649 &-0.828 &13.41 &20.2 & 3.450 &0.177 & $$ & $$ & $$ & $$  \\
                                   &0.06 & 8.989 & 6.648 &-1.291 &17.94 &17.9 & 3.238 &0.226 &1.95 & 1.77 & 1.00 & $$  & 0.15  \\
                                   &0.08 & 8.987 & 6.649 &-1.796 &26.36 &16.4 & 2.106 &0.280 & 1.94 & 1.75 & 1.00 & $$ & 0.20  \\
                                   &0.10 & 8.986 & 6.649 &-2.897 &33.21 &15.1 & 1.763 &0.284 & 2.02 & 1.72 & 1.00 & $$ & 0.22  \\
\end{tabular}
\end{ruledtabular}
\end{table*}

Powder x-ray diffraction (XRD) measurements at room temperature
were done to examine the phase purity. Figure 1 shows the typical
XRD patterns of SrZn$_x$Cu$_{2-x}$(BO$_3$)$_2$ and
SrCu$_2$(Si$_x$B$_{1-x}$O$_3$)$_2$. The diffraction peaks can be
well indexed on the known tetragonal unit cell with space group of
\emph{I}\={4}2m. It indicates that these samples are single phase
in the range of 0$\leq$$x$$\leq$0.15 for
SrZn$_x$Cu$_{2-x}$(BO$_3$)$_2$ and 0$\leq$$x$$\leq$0.1 for
SrCu$_2$(Si$_x$B$_{1-x}$O$_3$)$_2$. However, no pure samples can
be obtained if substitution content $x$ is further increased.
Table 1 shows the structure parameters obtained from the
least-squares calculation of 2$\theta$ values. Figure 2 shows the
relative change of lattice parameters as a function of $x$. It can
be seen that the in-plane parameter $a$ decreases monotonously
with increasing $x$. However, the $c$-axis parameter $c$ remains
almost unchanged. This is different from the out-of-plane
substitution where $a$ keeps almost unchanged but $c$ changes
monotonously with doping\cite{Liu1}. The induction-coupled plasma
(ICP) method was employed to determine the atomic ratio. The
experimental results listed in Table 1 are close to the nominal
composition. These results suggest that the substitution takes
place in the CuBO$_3$ plane.

\begin{figure}
\includegraphics[width=0.85\linewidth]{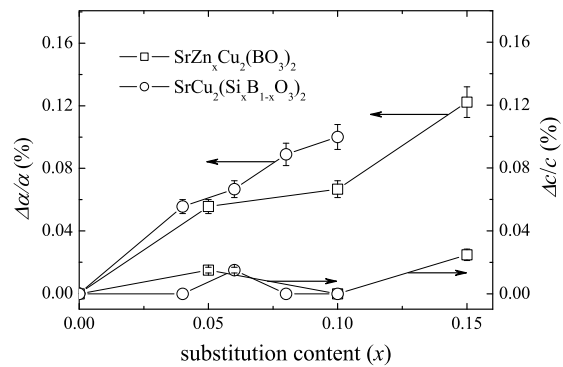}
\caption{\label{fig:fig_2} {Substituting content dependence of
lattice parameters for SrZn$_x$Cu$_{2-x}$(BO$_3$)$_2$ and
SrCu$_2$(Si$_x$B$_{1-x}$O$_3$)$_2$.}}
\end{figure}

The magnetization $M$ was measured using a commercial Quantum
Design PPMS. The resistivity at room temperature was measured with
a four-probe method using the Keithley low current source model
6487 Picoammeter.

\begin{figure}[t]
\includegraphics[width=0.75\linewidth]{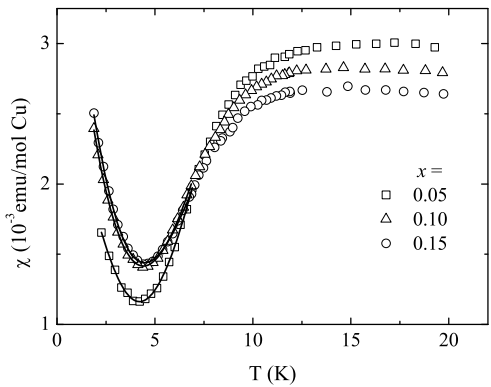}
\caption{\label{fig:fig_3} {Temperature dependence of the magnetic
susceptibility for SrZn$_x$Cu$_{2-x}$(BO$_3$)$_2$ in a field of
1.0 T, the solid lines are the fitting results with Eq.
(\ref{chi}).}}
\end{figure}

\begin{figure}[b]
\includegraphics[width=0.75\linewidth]{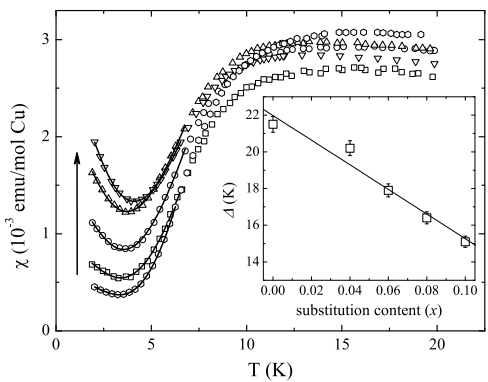}
\caption{\label{fig:fig_4}{Temperature dependence of the magnetic
susceptibility for SrCu$_2$(Si$_x$B$_{1-x}$O$_3$)$_2$ in a field
of 1.0 T, the solid lines are the fitting results with Eq.
(\ref{chi}), the arrow direction denotes the increase of $x$ from
0 to 0.10. The inset shows the spin gap $\Delta$ at different
substituting content $x$.}}
\end{figure}

Figure 3 shows the dc magnetic susceptibility $\chi$=$M/H$ as a
function of temperature $T$ for SrZn$_x$Cu$_{2-x}$(BO$_3$)$_2$
with $x$=0.05, 0.10, and 0.15 at 1.0 T from 2 to 20 K. For all
Zn-substituting samples, $\chi$ follows the Curie-Weiss law in
high temperature regime ($T$$>$20 K). At $\sim$10 K (defined as
$T_{SG}$), a spin gap opens and $\chi$ drops sharply. The
susceptibility behavior is typical for a low-dimensional spin-gap
system, and it is similar to our previous out-of-plane doping
result\cite{Liu1}. A Curie-like upturn contributed by magnetic
impurities\cite{Kageyama1, Zorko1} is observed below 4 K. Compared
with our previous Y-doping case\cite{Liu1}, the magnitude of
upturn here is more pronounced. The reason is some Cu$^{2+}$ were
replaced by spinless Zn$^{2+}$, which creates some individual
Cu$^{2+}$ with spin-1/2 by destroying a certain number of
Cu$^{2+}$-Cu$^{2+}$ dimers. As it is known that the contribution
of magnetic impurities to magnetic susceptibility follows the
Curie-Weiss law, $\chi$$\sim$$1/(T-\theta)$. However, the magnetic
susceptibility of anti-ferromagnetic dimers satisfies exponential
law, $\chi$$\sim$$e^{-1/T}$. Therefore, the contribution of these
magnetic impurities is very notable in low temperature range. The
upturn of $\chi$ at low temperatures rises with increasing $x$,
which indicates that more magnetic impurities were introduced.
These magnetic-impurity concentration can be reflected by
effective moment $\mu_\textbf{eff}$. From the temperature
dependence of $\chi$ at low temperatures, we can determine
$\mu_\textbf{eff}$
(\emph{C}$^\prime$=\emph{N}${\mu}$$_\textbf{eff}^2$/{3\emph{k}$_B$)
and the excitation gap $\Delta$ using the following
relation\cite{Liu1}:
\begin{equation}
{\chi}=\frac{C^\prime}{T-\theta^\prime}+a\exp(-\Delta/T)+\chi_{0}.
\label{chi}
\end{equation}
The first term represents the contribution of spin-$1/2$ magnetic
impurities. The second term represents the contribution from the
anti-ferromagnetic dimer structure. The third term is a small
constant which includes the diamagnetic contribution of both the
sample holder and the sample itself. The fitting results of
$\chi_0$, $C^{\prime}$, $\Delta$ as well as $\mu_\textbf{eff}$ are
given in Table 1. It can be seen that, with $x$ increased from 0
to 0.15, $\Delta$ is suppressed from 21.5 to 15 K, while
$\mu_\textbf{eff}$ increased from 0.110 to 0.212 $\mu_\textbf{B}$.

\begin{figure}[b]
\includegraphics[width=0.75\linewidth]{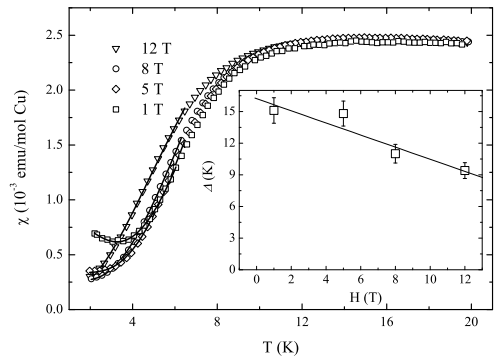}
\caption{\label{fig:fig_5} {Magnetic susceptibility $\chi$ as a
function of temperature $T$ of
SrCu$_2$(Si$_{0.1}$B$_{0.9}$O$_3$)$_2$ up to 12 T. The inset shows
the spin gap $\Delta$ as a function of fields $H$.}}
\end{figure}

Figure 4 shows a change of $\chi$ for
SrCu$_2$(Si$_x$B$_{1-x}$O$_3$)$_2$ plotted as a function of $T$.
We find $\chi$ has an analogous behavior as that of
Zn-substituting samples discussed above. With the same method as
treating Zn-substituting samples, $\Delta$ and $\mu_\textbf{eff}$
can be obtained for Si-substituting samples, which are listed in
Table 1. From the inset of Fig. 4, one can see that $\Delta$
reduces monotonously with $x$ increased. From the value of the
spin gap listed in Table 1, one can see that Zn- and
Si-substituting have a similar effect on the reduction of the spin
gap. It should be noted that the spin gap deduced from the
magnetic susceptibility data is $\sim$ 10 K less than that from
ESR\cite{Nojiri}, Raman scattering\cite{Lemmens}, neutron
scattering\cite{Kageyama3, Kakurai}, specific
heat\cite{Kageyama4}, and NMR\cite{Kodama} measurements. The
reason maybe comes from the upturn of susceptibility. However, it
does not affect the qualitative analysis of the effect of
substitution on the spin gap.

In order to further investigate the intrinsic nature of magnetic
state of in-plane substituted compound. We have measured
 $\chi(T)$ at different magnetic fields $H$ and $M(H)$ at different temperatures
$T$ for SrCu$_2$(Si$_{0.1}$B$_{0.9}$O$_3$)$_2$. Figure 5 shows
$\chi$ as a function of $T$ up to 12 T. The low-temperature
susceptibility upturn is gradually suppressed by rasing external
fields. From the inset of Fig. 5, a linear field dependence of
spin gap is observed, which could be explained by Zeeman
splitting. This result is also consistent with that derived from
specific heat data for out-of-plane doping cases\cite{Liu1}.
Figure 6 shows the $M(H)$ at different temperatures. At low
temperatures ($T$$<$$T_{SG}$), the initial linear field dependence
of $M(H)$ signifying that $M(H)$ is dominated by magnetic
impurity; with $H$ further increased, $M(H)$ gradually increases
with $H$, in this stage external field $H$ firstly saturates the
$M(H)$ of magnetic impurity and then breaks some
Cu$^{2+}$-Cu$^{2+}$ dimers. At higher temperatures
($T$$\geq$$T_{SG}$), the linear relationship between $M(H)$ and
$H$ can be interpreted as Cu$^{2+}$-Cu$^{2+}$ dimers were broken
by thermal fluctuation. We measured $\chi$ with zero-field-cooled
(ZFC) as well as field-cooled (FC), no spin-frozen or hysteresis
was found. These results suggest that there are two kinds of
magnetic systems in SrCu$_2$(Si$_{0.1}$B$_{0.9}$O$_3$)$_2$. One is
dimer system, and the other one is magnetic impurity system
induced by substitution.

The resistivity $\rho$ measurement results are listed in Table 1.
The resistivity decreases with increasing of substituting content
for both Cu- and B-site substitutions. However, we find that the
decrease in resistivity for in-plane substitution is much less
than that for out-of-plane substitution\cite{Liu1}, for example,
the resistivity for 10$\%$ La doping is about 2 orders less than
that for undoped SrCu$_{2}$(BO$_3$)$_2$. In case of Zn-substituting, no
carriers are introduced since
Zn$^{2+}$ ion has the same valence to that of Cu$^{2+}$. While for
Si-substituting case, $sp^2$ hybridization of B-O bond might not
be destroyed, therefore no additional conducting channels were
formed. The balance of electrovalence between Si$^{4+}$ and
B$^{3+}$ may be adjusted by oxygen content.

\begin{figure}
\includegraphics[width=0.75\linewidth]{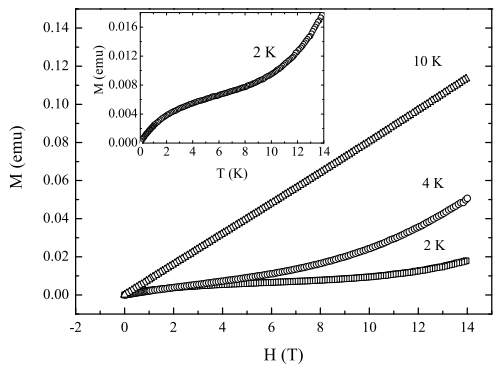}
\caption{\label{fig:fig_6} {Magnetization $M$ as a function of
magnetic fields $H$ of SrCu$_2$(Si$_{0.1}$B$_{0.9}$O$_3$)$_2$. The
inset shows the magnetization at 2 K.}}
\end{figure}

In conclusion, we have synthesized a series of in-plane
substituted materials, including Cu-site
(SrZn$_x$Cu$_{2-x}$(BO$_3$)$_2$) and B-site substitution
(SrCu$_2$(Si$_x$B$_{1-x}$O$_3$)$_2$). Their structures and
magnetic properties have been systematically studied. XRD analysis
suggests that these compounds are single-phase materials and their
in-plane lattice parameter depends systematically on the
substituting content $x$. The magnetism measurements indicate that
there are two kinds of magnetic systems in the substituted samples.
One is magnetic impurity system, and the other one is dimer structure
system. The spin gap can be reduced by element substitution. No
superconductivity appeared in the in-plane substituted samples.

This work is supported by the National Natural Science Foundation
of China.

\end{document}